\documentclass[a4paper, amsfonts, amssymb, amsmath, reprint, showkeys, footinbib, twoside,superscriptaddress,floatfix,longbibliography]{revtex4-1}

\usepackage{graphicx}%
\usepackage{dcolumn}%
\usepackage{bm}%
\usepackage{float}
\usepackage{xcolor}
\usepackage{mhchem}
\usepackage{gensymb}
\usepackage{verbatim}
\usepackage{newtxtext}
\usepackage[slantedGreek]{newtxmath}
\usepackage{amsmath}

\usepackage[T1]{fontenc}
\usepackage{hyphenat}
\usepackage{comment}
\usepackage[pdftex, bookmarks, pdffitwindow=false, pdfstartview=FitH, pdfdisplaydoctitle, colorlinks, plainpages=false, pdftitle={},pdfauthor={}, pdfpagelabels, hypertexnames, citecolor={blue!50!black},linkcolor={blue!50!black}, urlcolor={blue!50!black}, pdflang={en}, hyperfootnotes=false, breaklinks]{hyperref}

\usepackage{siunitx}
\usepackage{comment}

\DeclareUnicodeCharacter{0308}{HERE!HERE!} 

\DeclareSIUnit\angstrom{\text {Å}}

\makeatletter
\renewcommand{\@seccntformat}[1]{}
\makeatother

\makeatletter
\def\@bibdataout@aps{
 \immediate\write\@bibdataout{
 @CONTROL{
   apsrev41Control, author="48",editor="1",pages="0",title="0",year="1"
 }}
 \if@filesw
  \immediate\write\@auxout{\string\citation{apsrev41Control}}
 \fi
}
\makeatother

\begin{document}

\preprint{}

\title{Lattice distortion leads to glassy thermal transport in crystalline Cs$_3$Bi$_2$I$_6$Cl$_3$}

\author{Zezhu Zeng}
\email{zzeng@ist.ac.at}
\affiliation{The Institute of Science and Technology Austria, Am Campus 1, 3400 Klosterneuburg, Austria}

\author{Zheyong Fan}
\affiliation{College of Physical Science and Technology, Bohai University, Jinzhou 121013, China}

\author{Michele Simoncelli}
\affiliation{Theory of Condensed Matter Group of the Cavendish Laboratory, University of Cambridge, United Kingdom}
\affiliation{Department of Applied Physics and Applied Mathematics, Columbia University, New York (USA)}

\author{Chen Chen}
\email{ccmldn@gbu.edu.cn}
\affiliation{School of Physical Sciences, Great Bay University, Dongguan, Guangdong, China}

\author{Ting Liang}
\affiliation{Department of Electronic Engineering and Materials Science and Technology Research Center, The Chinese University of Hong Kong}

\author{Yue Chen}
\affiliation{Department of Mechanical Engineering, The University of Hong Kong, Pokfulam Road, Hong Kong SAR, China}

\author{Geoff Thornton}
\affiliation{London Centre for Nanotechnology, University College London, London WC1H 0AJ, United Kingdom}

\author{Bingqing Cheng}
\email{bingqingcheng@berkeley.edu}
\affiliation{Department of Chemistry, University of California, Berkeley, CA, USA}
\affiliation{The Institute of Science and Technology Austria, Am Campus 1, 3400 Klosterneuburg, Austria}

\date{\today}

\begin{abstract}

The glassy thermal conductivities observed in crystalline inorganic perovskites such as Cs$_3$Bi$_2$I$_6$Cl$_3$ 
is perplexing and lacking theoretical explanations.
Here, we first experimentally measure such its thermal transport behavior from 20~K to 300~K, after synthesizing Cs$_3$Bi$_2$I$_6$Cl$_3$ single crystals.
Using path-integral molecular dynamics simulations driven by machine learning potentials, we reveal that Cs$_3$Bi$_2$I$_6$Cl$_3$ has large lattice distortions at low temperatures,
which may be related to the large atomic size mismatch.
Employing the Wigner formulation of thermal transport, we reproduce the experimental thermal conductivities based on lattice-distorted structures.
This study thus provides a framework for predicting and understanding glassy thermal transport in materials with strong lattice disorder.

\end{abstract}

\maketitle

\begin{figure}[htb]
    \centering
    \includegraphics[scale=1.0]{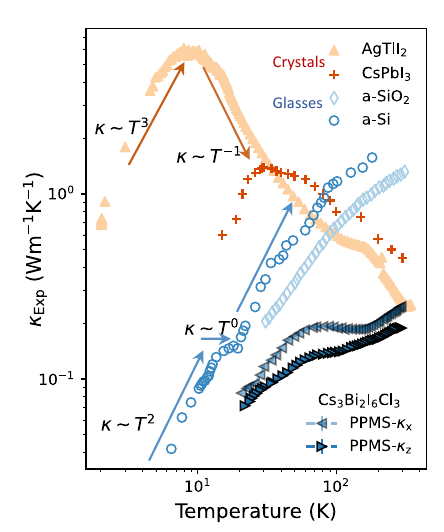}
    \caption{Experimental lattice thermal conductivity ($\kappa$) for AgTlI$_2$ \cite{zeng2024pushing} and CsPbI$_3$ \cite{lee2017ultralow} crystals, amorphous Si \cite{zink2006thermal} and SiO$_2$ \cite{cahill1990thermal} glasses,
    and single crystal Cs$_3$Bi$_2$I$_6$Cl$_3$ measured by us for both parallel ($\kappa_{z}$) and perpendicular ($\kappa_{x} = \kappa_{y}$) directions relative to the Bridgman growth direction.}
    \label{fig:fig1}
\end{figure}

A foundational theory for the lattice thermal conductivity ($\kappa$) in crystals was proposed by Peierls in 1929~\cite{peierls1929kinetischen} based on the phonon gas picture:
$\kappa$ has a $T^3$ behavior at low temperatures, followed by a decrease with a $T^{-1}$ dependence due to Umklapp phonon scattering.
This behavior is observed in many materials, such as thermoelectric AgTlI$_2$ \cite{zeng2024pushing} and inorganic perovskite CsPbI$_3$ \cite{lee2017ultralow},
as illustrated in Fig.~\ref{fig:fig1}.

For glasses, such as SiO$_2$ \cite{cahill1990thermal} and B$_2$O$_3$ \cite{freeman_thermal_1986},
the behavior of $\kappa$ is distinct (see Fig.~\ref{fig:fig1}):
At low temperatures ($T \lesssim 1$~K), $\kappa(T) \sim T^2$, which 
may be explained by quantum tunneling~\cite{anderson1972anomalous};
between $5$~K and $25$~K, $\kappa(T)$ plateaus, which may be related to the boson peak of glasses~\cite{schirmacher2007acoustic};
At higher $T$, $\kappa$ increases with temperature,
which can be explained by the Allen-Feldman (AF) model~\cite{allen1989thermal, allen1993thermal}
based on couplings between quasi-degenerate atomic vibrational modes.

However, $\kappa$ of some crystalline materials do not follow the typical behavior of crystals but are rather glassy, e.g. layered perovskite Cs$_3$Bi$_2$I$_6$Cl$_3$ \cite{acharyya2022glassy}, quasi-one-dimensional perovskite BaTiS$_3$ \cite{sun2020high,zhao2022orientation}, and Ruddlesden-Popper perovskites Ba$_3$Zr$_2$S$_7$ and Ba$_4$Zr$_2$S$_{10}$ \cite{hoque2023ruddlesden}. 
In this study, we synthesized a high-quality single crystal (see Methods section for details) of Cs$_3$Bi$_2$I$_6$Cl$_3$ (space group $P\overline{3}m1$) and measured its $\kappa$ from 20~K to 300~K (see Fig. \ref{fig:fig1}) . 
Interestingly, both $\kappa_{x}$ and $\kappa_{z}$ in Cs$_3$Bi$_2$I$_6$Cl$_3$ deviate from the expected $T^3$ dependence at low $T$ and lack crystalline-like peak in the entire temperature range. 
In addition, $\kappa_{x}$ exhibits an unusual wide plateau region (60–150~K), while $\kappa_{z}$ steadily increases with increasing temperature.  

Some other materials also exhibit measured glassy $\kappa$ below room temperature.
For example, (KBr)$_{1-x}$(KCN)$_{x}$ \cite{cahill1992lower,beekman2017inorganic} transitions from crystalline to glassy $\kappa$ with an increase in the molar fraction of KCN, which was attributed to binary doping. 
The glassy $\kappa$ in clathrates Sr$_8$Ga$_{16}$Ge$_{30}$ and Eu$_8$Ga$_{16}$Ge$_{30}$ \cite{sales2001structural} was rationalized by the rattling atoms within cage-like structures. 
In contrast, Cs$_3$Bi$_2$I$_6$Cl$_3$ lacks the binary doping as mixed crystals or a cage-like structure, suggesting a distinct mechanism behind its glassy $\kappa$. 
Acharyya \textit{et al.} \cite{acharyya2022glassy} first reported the glassy $\kappa$ of crystalline Cs$_3$Bi$_2$I$_6$Cl$_3$, and rationalized the slow increase of $\kappa$ near room temperature based on local atomic disorder and low sound velocity, but the mechanism governing the glassy $\kappa$ at lower temperatures remains unclear.

It is challenging to calculate $\kappa$ for solids with strong anharmonicity at low temperatures:
Regular molecular dynamics (MD) does not consider nuclear quantum effects (NQE).
Quantum MD methods, such as path integral MD (PIMD) \cite{parrinello1984jcp}, centroid MD (CMD) \cite{cao1994formulation}, ring polymer MD (RPMD) \cite{habershon2013ring}, and thermostatted ring polymer MD (TRPMD) \cite{rossi2014remove}, can incorporate the quantum delocalization of nuclei and quantum thermodynamic properties, while they cannot address real-time dynamic properties and thus require approximations for the non-linear heat flux operator~\cite{luo2020capturing,sutherland2021nuclear}.
There is still debate on whether these quantum MD methods are accurate for computing thermal conductivities: Luo et al. \cite{luo2020capturing} reported that RPMD combined with the Green-Kubo theory could reproduce the $\kappa$ of ice Ih from 150~K to 200~K.
Moreover, they argued that there is no essential difference between CMD and RPMD in computing quantum thermal conductivity, while TRPMD slightly underestimates $\kappa$ of ice. 
However, Manolopoulos et al. \cite{sutherland2021nuclear} observed significant overestimation of Green-Kubo $\kappa$ using CMD for liquid para-hydrogen below 30~K. They attributed these errors to the classical heat capacities used in CMD and proposed a quantum heat capacity scaling to achieve better agreement with experiments.

On the other hand, recent advancements~\cite{caldarelli2022many} in lattice dynamics such as Wigner transport equation (WTE)~\cite{simoncelli2019unified,simoncelli2022wigner,caldarelli2022many} and quasi harmonic Green-Kubo (QHGK) theory~\cite{isaeva2019modeling,fiorentino_green-kubo_2023}, by accounting for both (intraband) propagation ($\kappa_{\rm P}$) and (interband, Zener-like tunneling) coherence ($\kappa_{\rm C}$) 
thermal conductivities, offer new methods to compute $\kappa$ of complex materials accounting for the Bose-Einstein statistics of vibrations, anharmonicity and disorder. 
For crystals, both WTE and QHGK can accurately compute the bulk limit of $\kappa$ by integrating microscopic, mode-resolved vibrational properties over the entire Brillouin zone (BZ), and example applications include WTE for perovskite CsPbBr$_3$ \cite{simoncelli2019unified} and QHGK for diamond-cubic silicon \cite{barbalinardo2020efficient}. 
For glasses, calculating the bulk limit of the conductivity requires ensuring that the finite-size atomistic models are large enough to realistically describe amorphicity. Various methods have been developed to address this challenge \cite{feldman_thermal_1993,simoncelli2023thermal,fiorentino2023hydrodynamic,harper2024vibrational}. These methods share the key feature that when glassy atomistic models are sufficiently large to capture disorder, the conductivity remains unchanged upon further increasing the model size or applying different boundary conditions at the simulation-domain boundaries
\footnote{A recent work \cite{simoncelli2024temperature} has shown that in silica-based materials where anharmonic effects are large enough to be accurately described by the common relaxation-time approximation (like those reported here) simulations containing a few thousand atoms are sufficiently large to describe the opposite scaling laws of the conductivities of crystals and glasses upon heating. Specifically, simulations containing 6804 atoms are large enough to describe the propagation-dominated and thermally damped conductivity of AV tridymite crystal, as well as the tunneling-dominated and thermally activated conductivity of TOGA silica.}.
In practice, to verify this, one must check that the conductivity remains unchanged when predicted using increasingly larger cells with standard periodic boundary conditions (PBC); an additional test is to verify that the PBC $\kappa$ is equivalent to $\kappa$ computed by averaging over many different boundary conditions \cite{simoncelli2023thermal,harper2024vibrational}. The former test practically corresponds to performing calculations at $\boldsymbol{q}=\boldsymbol{0}$ only in models with increasingly larger size, while the latter corresponds to finding unchanged $\kappa$ when comparing a calculation at $\boldsymbol{q}=\boldsymbol{0}$ only with one performed by averaging over wavevectors $\boldsymbol{q}\neq\boldsymbol{0}$ belonging to the  small BZ of the disordered model.


\begin{figure*}[htb]
    \centering
    \includegraphics[scale=0.85]{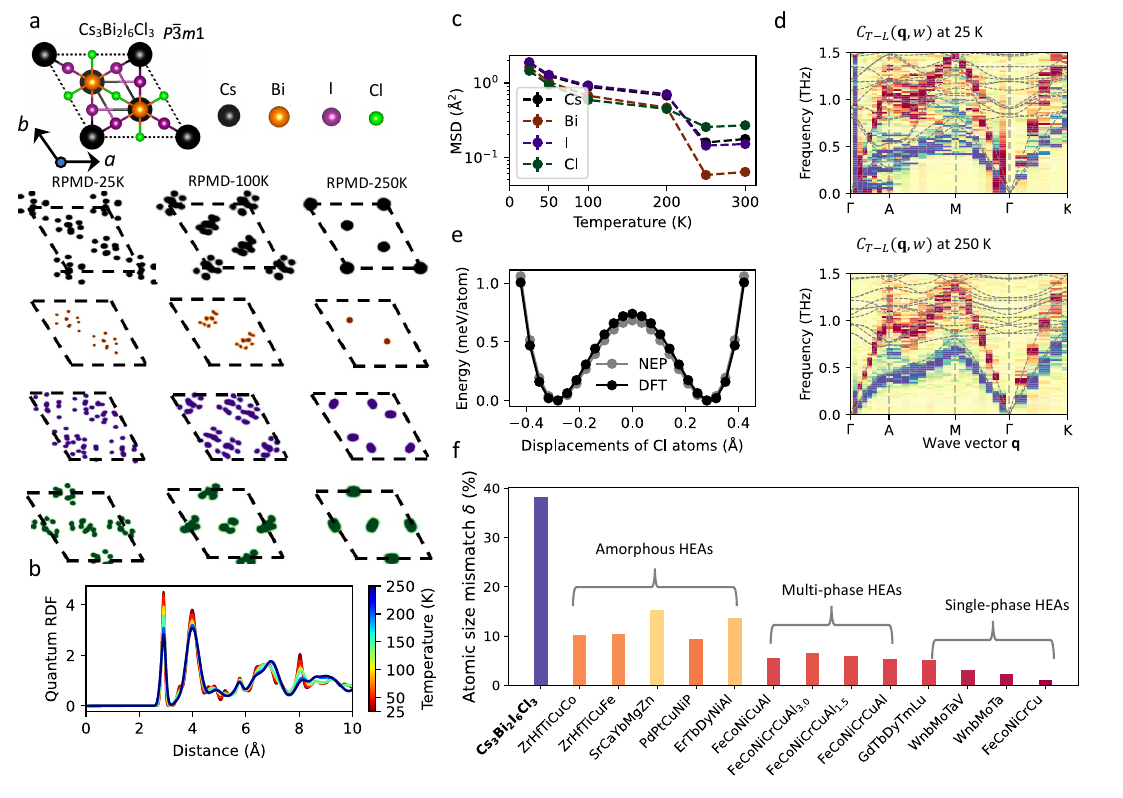}
    \caption{\textbf{a} Atomic probability distribution of four elements in Cs$_3$Bi$_2$I$_6$Cl$_3$ from NEP ring-polymer molecular dynamics (RPMD) simulations with a 5 $\times$ 5 $\times$ 10 supercell (3500 atoms) at 25~K, 100~K, and 250~K. 
    \textbf{b} Calculated radial distribution function (RDF) of Cs$_3$Bi$_2$I$_6$Cl$_3$ at different temperatures based on the RPMD simulations with a 678-atoms supercell.
    \textbf{c} Mean square displacements (MSD) from 25~K to 300~K computed from NEP ring-polymer molecular dynamics (RPMD) simulations with a 5 $\times$ 5 $\times$ 10 supercell (3500 atoms), using the $P\overline{3}m1$ crystalline Cs$_3$Bi$_2$I$_6$Cl$_3$ as the reference structure.  
    \textbf{d} Transverse (red) and longitudinal (blue) current correlation functions $C_{T-L}(\mathbf{q}, \omega)$ as a function of wave vector $\mathbf{q}$ and frequency $\omega$, calculated from RPMD trajectories at 25~K and 250~K with a 16 $\times$ 16 $\times$ 16~supercell (57344 atoms). The dashed curves show phonon dispersions of Cs$_3$Bi$_2$I$_6$Cl$_3$, computed using the temperature-dependent effective potential (TDEP) approach with the $P\overline{3}m1$ crystal structure.
    \textbf{e} Potential energy surfaces of the soft mode at $\Gamma$ point calculated using NEP and DFT by displacing Cl atoms in the unit cell. 
    \textbf{f} Atomic size mismatch parameter ($\delta$) for crystalline Cs$_3$Bi$_2$I$_6$Cl$_3$ and selected high-entropy alloys (HEAs)~\cite{he2018lattice} with single-phase, multiple-phase, and amorphous-phase structures.} 
    \label{fig:fig2}
\end{figure*}

\subsection{Lattice distortion at low temperatures}
We first constructed a machine-learning-based neuroevolution potential (NEP) \cite{fan2021neuroevolution} version 4 \cite{Song2024nc}, for Cs$_3$Bi$_2$I$_6$Cl$_3$ across a broad range of temperature and strain,
based on the PBEsol functional \cite{perdew2008restoring} with the D3 dispersion correction \cite{grimme2011effect}.
We then simulated the crystal structure of Cs$_3$Bi$_2$I$_6$Cl$_3$ between 25~K and 300~K.
A 5 $\times$ 5 $\times$ 10 supercell with 3,500 atoms was used to perform NEP PIMD simulations. 
Fig. \ref{fig:fig2}a shows the projection of atom positions onto the two-dimensional $a$-$b$ plane in a single unit cell with 10-fold replication (1 $\times$ 1 $\times$ 10) along the $z$ direction.
At 250~K, all atoms vibrate near their equilibrium lattice positions in the $P\overline{3}m1$ crystal structure, consistent with results of the crystal structure characterization from our single-crystal XRD experiment at 300~K (Table S2 of Supplementary Information (SI)). 
In contrast, at 25~K
the atomic position distributions have multiple peaks.
At 100~K, the peaks grow in size but become closer together, suggesting a reduction in lattice distortion but with stronger thermal vibrations. 
To the best of our knowledge, this distortion and related disorder-to-order transition is reported for the first time in this work. 
This distortion also exhibits a special correlation. Specifically, we selected a neighboring cell adjacent to the reference cell (shown in Fig. \ref{fig:fig2}a) and performed the same analysis of the atomic probability distribution along the $a$-$b$ plane (see Fig. S4). The results reveal that the projection patterns of the two cells are remarkably similar. 
Moreover, the distortion pattern also depends on the supercell size used to capture the disorder. Using a larger supercell (7 $\times$ 7 $\times$ 10) for RPMD simulations at 25~K revealed a distinct distortion pattern (Fig. S5). 
Larger supercells accommodate more commensurate soft modes, enhancing disorder through collective coupling (as discussed later).

Note that the scattered spots at 25~K and 100~K still cluster around the $P\overline{3}m1$ lattice sites, thus the calculated radial distribution functions (RDF) in Fig. \ref{fig:fig2}b based on RPMD simulations show characteristic peaks associated with crystals across the whole temperature range.
We compared the RDF (see Fig. S6) at 25~K, 100~K, and 250~K computed from RPMD and classical MD simulations. The RDF at 25~K exhibits a distinct difference, with the RDF from the RPMD simulations showing a broader first peak. This result indicates that, despite all atoms in Cs$_3$Bi$_2$I$_6$Cl$_3$ being relatively heavy, PIMD-based simulations are necessary to accurately assess the structural and thermodynamic properties of the system.

To further quantify the lattice distortion, we computed
the mean square displacement (MSD) of the four elements based on RPMD trajectories, using the $P\overline{3}m1$ structure as a reference.
As shown in Fig. \ref{fig:fig2}c, MSD decreases with increasing temperature, with
a notable drop between 200~K and 250~K that corresponds to the structure transition from disorder to order.
Above 250~K, as the crystal structure reverts to the $P\overline{3}m1$ phase, MSD values increase with temperature, which is a typical tendency in ordered crystals. 
We also compared the MSD obtained from RPMD and classical MD simulations (Fig. S7). 
The results highlight the importance of PIMD-based simulations in capturing accurate atomic vibrations at temperatures below 100~K, as the MSD from RPMD is significantly higher than that from classical MD at 25~K and 50~K.

The strong disorder in crystalline Cs$_3$Bi$_2$I$_6$Cl$_3$ at low temperatures is unusual for two reasons: 
First, in many disordered crystals, only a part of elements exhibit spatially correlated disorder~ \cite{keen2015crystallography,liang2023structural}, where lattice distortions follow a specific pattern with spatial correlations, e.g.
in thermoelectric PbTe the Pb atoms exhibit large deviation along specific $\langle$100$\rangle$ directions~\cite{bovzin2010entropically},
in metal-organic frameworks~\cite{meekel2021correlated} the correlated distribution of linker vacancies leads to a patterned disordering.
In contrast, all elements in Cs$_3$Bi$_2$I$_6$Cl$_3$ have disorder.
Second, the other disordered crystals are typically ordered at low $T$ and become disordered at high $T$ ~\cite{keen2015crystallography}, while Cs$_3$Bi$_2$I$_6$Cl$_3$ is opposite.

Strong lattice distortion can lead to substantial phonon-disorder scatterings. 
To investigate this, we used a 16 $\times$ 16 $\times$ 16~supercell with 57344~atoms to compute the longitudinal and transverse current correlation functions based on RPMD trajectories at 25~K and 250~K (as shown in Fig. \ref{fig:fig2}d with heatmaps), which incorporates both full-order lattice anharmonicity and NQE.
The longitudinal and transverse current correlation functions \cite{fransson2021dynasor}, $C_L(\mathbf{q}, t)$ and $C_T(\mathbf{q}, t)$ are
\begin{equation}
    \begin{aligned}
     & C_L(\mathbf{q}, t)=\frac{1}{N}\left\langle\mathbf{j}_L(\mathbf{q}, t) \cdot \mathbf{j}_L(-\mathbf{q}, 0)\right\rangle, \\
     & C_T(\mathbf{q}, t)=\frac{1}{N}\left\langle\mathbf{j}_T(\mathbf{q}, t) \cdot \mathbf{j}_T(-\mathbf{q}, 0)\right\rangle,
    \end{aligned}
\end{equation}
where $\mathbf{j}_L(\mathbf{q}, t)$ and $\mathbf{j}_T(\mathbf{q}, t)$ are given by
\begin{equation}
    \begin{aligned}
     & \mathbf{j}_L(\mathbf{q}, t) = \sum_i^N \left( \mathbf{v}_{\mathbf{i}}(t) \cdot \hat{\mathbf{q}} \right) \hat{\mathbf{q}} \mathrm{e}^{\mathrm{i} \mathbf{q} \cdot \mathbf{r}_i(t)}, \\
     & \mathbf{j}_T(\mathbf{q}, t) = \sum_i^N \left[\mathbf{v}_{\mathbf{i}}(t) - \left( \mathbf{v}_{\mathbf{i}}(t) \cdot \hat{\mathbf{q}} \right) \hat{\mathbf{q}}\right] \mathrm{e}^{\mathrm{i} \mathbf{q} \cdot \mathbf{r}_{\mathbf{i}}(t)},
    \end{aligned}
\end{equation}
with $N$, $\mathbf{q}$, $\mathbf{r}_i$, and $\mathbf{v}_i$ representing the total number of atoms, wave vector, atomic position vector, and velocity vector, respectively.
For comparison, Fig. \ref{fig:fig2}d also shows the renormalized phonon dispersions at 25~K and 250~K using gray lines, calculated using the temperature-dependent effective potential (TDEP) \cite{hellman2013temperature} based on the $P\overline{3}m1$ structure. 
TDEP incorporates temperature-dependent phonon frequency correction arising from lattice anharmonicity beyond the third order. 
At 250 K, the phonon modes are well-defined and all real, with frequencies from current correlation functions closely matching those derived from TDEP calculations, 
which reveals that higher-than-third-order anharmonic renormalization to harmonic phonon frequencies (see Fig. S18) at 0 K.
In contrast, phonon scattering at 25~K is significantly more severe than at 250~K, especially along the wave vector paths from $\Gamma$ to A and from M to $\Gamma$ in the first BZ, 
indicating substantial phonon-disorder interactions.
Additionally, we observe unusual overdamped modes near the BZ center at 25~K, as similarly observed in perovskite CsPbBr$_3$ at 385~K in the tetragonal phase and 500~K in the cubic phase\cite{lanigan2021two}, 
while no imaginary modes are observed in TDEP results.
This comparison suggests that lattice dynamics based on ordered crystalline Cs$_3$Bi$_2$I$_6$Cl$_3$ is not suitable for comprehending vibrational modes and heat transport of Cs$_3$Bi$_2$I$_6$Cl$_3$ at low temperatures.

To rationalize the large lattice distortion, we computed potential energy surfaces (PES) of the imaginary phonon modes at high-symmetry points by displacing corresponding atoms along their eigenvectors.
We found shallow and double-well PES shown in Fig. \ref{fig:fig2}e (for $\Gamma$ point) and Fig. S18 (for other high-symmetry points),
leading to significant atomic displacements towards two local minima with large distances of about~0.7~\AA.
In Fig. \ref{fig:fig2}e, the energy barrier along the imaginary mode of $\Gamma$ point is only 0.7~meV/atom, which is about the energy scale of thermal fluctuations at about 10~K.
The shallow double-well PES associated with imaginary modes from the low-frequency branch along the $\Gamma$-A-M-$\Gamma$-K path (Fig. S18) collectively drive lattice distortion in Cs$_3$Bi$_2$I$_6$Cl$_3$. This branch collapse typically correlates with overall structural disorder, rather than an isolated soft mode driving a phase transition to an ordered phase with lower symmetry.
After the distortion, the atoms in the distorted structure exhibit local vibrations with negligible MSD (see Fig. S8) around their new equilibrium positions, without any observable hopping behavior. The calculated harmonic vibrational density of states (see Fig. S9) for these distorted structures further confirms the absence of imaginary frequencies.

Lattice distortions in crystals are often attributed to atomic size mismatch~\cite{zhang2008solid}, i.e.
\begin{equation}
    \delta \%=100 \% \sqrt{\sum_{i=1}^n c_i\left(1-\frac{r_i}{\sum_{j=1}^n c_j r_j}\right)^2},
\end{equation}
where $c_i$ and $r_i$ denote the atomic fraction and atomic radius of the $i$th element, respectively. 
High-entropy alloys (HEAs) with large $\delta$ often have strong lattice distortions~\cite{song2017local}.
In comparison, 
the $\delta$ of Cs$_3$Bi$_2$I$_6$Cl$_3$ is larger than that of typical HEAs,
as shown in Fig. \ref{fig:fig2}f. 
The $\delta$ mainly come from the large difference in atomic radii of Cs and Cl.
The large $\delta$ not only shed light on the large lattice distortions, but also help explain why the $\kappa$ of Cs$_3$Bi$_2$I$_9$ exhibits typical crystalline behavior~\cite{acharyya2023extended} while
the $\kappa$ of Cs$_3$Bi$_2$I$_6$Cl$_3$ is glassy.

\subsection*{Thermal conductivity follows a glass-like mechanism}

We computed the $\kappa$ of Cs$_3$Bi$_2$I$_6$Cl$_3$ using both molecular dynamics and lattice dynamics. 
Classical MD simulations with Green-Kubo (GK) theory offer a straightforward method to compute the $\kappa$ of disordered crystals, as atomic disorder is inherently included.
At $T$ above 150~K, the computed $\kappa$ (see Fig. S15) agree well with our experimental data. 
However, below 150~K, the calculations show a distinct overestimation, due to the lack of NQEs in classical MD. 
To include NQEs, we also performed TRPMD simulations with GK theory.
The $\kappa$ calculated from TRPMD (see Fig. S15) does not show considerable differences from those computed using classical MD (i.e., still overestimating the experimental $\kappa$ at $T$ below 150~K), 
indicating potential limitations of the GK method within TRPMD for accurately assessing $\kappa$ at low $T$.
One possible limitation arises from the ambiguous definition of the non-linear heat flux operator in PIMD-based simulations~\cite{luo2020capturing, sutherland2021nuclear}.
In addition, any PIMD- or RPMD-based techniques are designed to sample quantum-mechanical distributions and cannot unambiguously capture real-time dynamics, limiting their ability to accurately compute heat flux.
We stress that the origin of the discrepancy between GK thermal conductivity predictions from classical or PIMD-based simulations and experimental results at extremely low temperatures remains unclear and warrants further investigations. 

\begin{figure}[htb]
    \centering
    \includegraphics[scale=0.56]{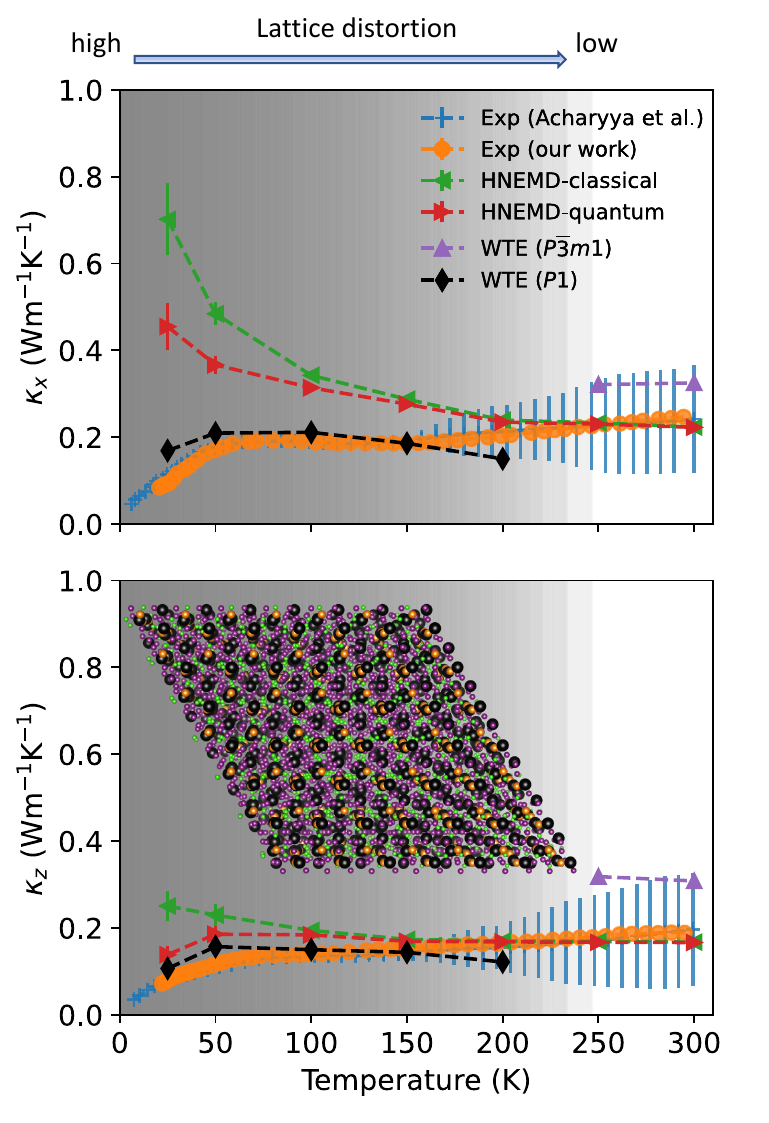}
    \caption{Calculated $\kappa$ for single crystal Cs$_3$Bi$_2$I$_6$Cl$_3$.
    Green and red triangles show results from homogeneous nonequilibrium MD (HNEMD) simulations, with and without empirical quantum corrections.
    At 250~K and 300~K, the purple triangles show the $\kappa$ calculated using the Wigner Transport Equation (WTE) with $P\overline{3}m1$ structure with Fourier interpolation.
    At $T$ between 25~K and 200~K, the diamonds show the WTE $\kappa$ on top of the relaxed distorted structures (4116 atoms) computed at only $\Gamma$ point (center of the Brillouin zone). 
    A relaxed disordered structure at 25~K is shown in the inset.  
    Additionally, the circle and cross symbols show experimental $\kappa$ from our measurements and from Ref.~\cite{acharyya2022glassy}, respectively.}
    \label{fig:fig3}
\end{figure}

Classical homogeneous non-equilibrium MD (HNEMD) simulations \cite{fan2019homogeneous,evans1982homogeneous}, with an empirical quantum correction~\cite{wang2023quantum} (detailed in Methods),
provide an alternative way to assess $\kappa$ without defining the heat flux in the PIMD-based framework.
The $\kappa$ from HNEMD are shown in Fig. \ref{fig:fig3}.
Our classical and quantum-corrected HNEMD results show good agreement with the experimental data above 
150~K.
Classical HNEMD $\kappa$ show an overestimation along both the $x$ and $z$ directions at $T$ below 150~K compared to the measurements, similar to the observation from the GK results with classical MD.
After applying the harmonic quantum correction~\cite{wang2023quantum} accounting for the modal heat capacity, the corrected $\kappa_{z}$ agrees well with the experimental data. 
However, $\kappa_{x}$ remains overestimated and fails to reproduce the glassy behavior.
This may be due to the limited scope of the harmonic quantum correction, which considers only the empirical correction of quantum effects on the heat capacity, but does not address mode-mode occupations that can affect phonon coupling.

As the MD based methods are not able to reproduce the experimental glassy $\kappa$ of Cs$_3$Bi$_2$I$_6$Cl$_3$, we resorted to lattice dynamics, 
which can incorporate the Bose-Einstein statistics of atomic vibration, also structure disorder \cite{harper2024vibrational} and anharmonicity \cite{simoncelli2022wigner}. 
For disordered Cs$_3$Bi$_2$I$_6$Cl$_3$ below 200~K, the lack of periodicity results in a thermodynamically large primitive cell, restricting WTE to contribute $\kappa$ from $\mathbf{q=0}$. 
Therefore, we employed the WTE \cite{simoncelli2019unified} at $\mathbf{q=0}$ to compute the $\kappa$ of disordered Cs$_3$Bi$_2$I$_6$Cl$_3$ below 200 K:
\begin{equation}
    \label{eq:wte}
    \begin{aligned}
        \kappa = & \frac{1}{\mathcal{V} N_c} \sum_{\mathbf{q}, s, s^{\prime}} \frac{\omega(\mathbf{q})_s+\omega(\mathbf{q})_{s^{\prime}}}{4}\left(\frac{C(\mathbf{q})_s}{\omega(\mathbf{q})_s}+\frac{C(\mathbf{q})_{s^{\prime}}}{\omega(\mathbf{q})_{s^{\prime}}}\right) \frac{\left\|\mathbf{v}(\mathbf{q})_{s, s^{\prime}}\right\|^2}{3} \\
        & \times \pi \mathcal{F}_{\left[\Gamma(\mathbf{q})_s+\Gamma(\mathbf{q})_{s^{\prime}}\right]}\left(\omega(\mathbf{q})_s-\omega(\mathbf{q})_{s^{\prime}}\right),
    \end{aligned}
\end{equation}
where $\mathcal{V}$ is the volume of the unit cell, $N_c$ is the number of $\mathbf{q}$-points in the summation, $s$ and $s^{\prime}$ are band indices at the wave vector $\mathbf{q}$ (with $\kappa_{\rm P}$ derived when $s=s^{\prime}$ and $\kappa_{\rm C}$ derived when $s \neq s^{\prime}$), $\omega$ is the vibrational frequency, $C(\mathbf{q})_s$ is the specific heat, and $\mathbf{v}(\mathbf{q})_{s, s^{\prime}}$ is the velocity operator. 
$\Gamma(\mathbf{q})_s$ is the anharmonic linewidth, and 
the Lorentzian distribution $\mathcal{F}$, having a full width at half maximum (FWHM) equal to $\Gamma(\mathbf{q})_s+\Gamma(\mathbf{q})_{s^{\prime}}$, is defined as:
\begin{multline}
    \label{eq:lw}
    \mathcal{F}_{\left[\Gamma(\mathbf{q})_s+\Gamma(\mathbf{q})_{s^{\prime}}\right]}\left(\omega(\mathbf{q})_s-\omega(\mathbf{q})_{s^{\prime}}\right) = \\
    \frac{1}{\pi} \frac{\frac{1}{2}\left(\Gamma(\mathbf{q})_s+\Gamma(\mathbf{q})_{s^{\prime}}\right)}{\left(\omega(\mathbf{q})_s-\omega(\mathbf{q})_{s^{\prime}}\right)^2+\frac{1}{4}\left(\Gamma(\mathbf{q})_s+\Gamma(\mathbf{q})_{s^{\prime}}\right)^2}.
\end{multline} 
Ideally, an infinitely large supercell is required to fully capture lattice disorder. Practically, we approximated this by using relaxed distorted supercells of 252, 896, 1750, 3024 and 4116 atoms to compute WTE $\kappa$ at $\mathbf{q=0}$. 
The relaxed structures, classified as $P1$ symmetry (see Fig. \ref{fig:fig3}), were used as the initial equilibrium configuration to compute group velocities and frequencies (see Methods for details). 
Given the computational expense of calculating anharmonic linewidths for large unit cells, we employed the widely used approximation that determines the  linewidths as a function of frequency (see Methods, as well as Fig. S19 and Tabel S3 of SI).

The WTE thermal conductivities exhibit variations with supercell size (Fig. S11), reflecting the combined influence of finite-size effects and atomic disorder.
Increasing the simulation cell's size yields a reduction of finite-size effects. 
Notably, the thermal conductivities computed for 3024-atom and 4116-atom supercells show relatively small differences. 
In Fig. \ref{fig:fig3}, we present the WTE results using the 4116-atom supercell. We observe that WTE thermal conductivities along both the $x$ and $z$ directions decrease with temperature below 50~K, contrasting with classical HNEMD and GK results but aligning more closely with experimental observations. Furthermore, from 25~K to 200~K, WTE results agree well with experiments within the uncertainty range. 
This comparison suggests that WTE, applied to distorted structures, effectively reproduces the glass-like thermal conductivities observed experimentally, particularly at extremely low temperatures.

We emphasize that the glass-like $\kappa$ of Cs$_3$Bi$_2$I$_6$Cl$_3$ below 200~K is mainly determined by structural disorder. 
This conclusion is further supported by an anharmonicity sensitivity analysis (see Fig.~S20), which shows that artificially doubling or halving the intrinsic linewidths yields unimportant changes in both 
$\kappa_{\rm x}$ and $\kappa_{\rm z}$. 
This behavior is fundamentally different from that observed in simple crystals having $\kappa$ limited by anharmonicity \cite{simoncelli2024temperature}, where doubling (halving) the linewidth directly halves (doubles) the conductivity. 
These findings confirm that in Cs$_3$Bi$_2$I$_6$Cl$_3$ below 200~K, structural disorder dominates over anharmonicity in limiting heat conduction, resulting in a conductivity displaying the glass-like  temperature dependence observed in experiments.

For the ordered $P\overline{3}m1$ phase at high $T$ (250~K and 300~K), the WTE predicts the bulk limit of $\kappa$ by integrating phonon properties over the entire BZ of the crystal \cite{simoncelli2019unified}.
We used this structure in the WTE calculation, which accounts for thermal expansion, renormalized phonon frequencies using TDEP \cite{hellman2013temperature}, and four-phonon interactions \cite{feng2016quantum,feng2017four}, essential corrections to accurately treat perovskites with strong lattice anharmonicity \cite{li2023wavelike}.
The calculated values of $\kappa$ are consistent with the experimental observations as shown in Fig. \ref{fig:fig3}. Further insights into thermal transport are obtained from the spectral-$\kappa$ at 300~K, as shown in Fig. S16. 
Notably, $\kappa_{\rm P}$ exhibits an extremely low value of $\sim$ 0.18~W/mK at 300~K along both $x$ and $z$ directions, and $\kappa_{\rm C}$ along the $z$-axis is almost equal to $\kappa_{\rm P}$ along the $z$-axis, implying considerable wavelike-tunneling heat transport in Cs$_3$Bi$_2$I$_6$Cl$_3$ at room temperature. 
Note that the experimental $\kappa$ above 200 K exhibits a subtle increase with temperature, which apparently deviates from the conventional $T^{-1}$ trend. 
This slight increase of $\kappa$ in crystalline Cs$_3$Bi$_2$I$_6$Cl$_3$ above 200~K can be attributed to the contribution from the coherence transport channel, which contribute to an additional wave-like $\kappa_{\rm C}$, as observed in complex crystals such as Ag$_8$GeSe$_6$ \cite{bernges2023analytical}.

Finally, we rule out two other possible factors that may affect the $\kappa$ comparisons between calculations and measurements.
To examine phonon-boundary scattering effects, we computed the spectral mean free path (MFP) of Cs$_3$Bi$_2$I$_6$Cl$_3$ at 25 K. 
At 25 K, the maximum MFP (Fig. S17) along the $x$ direction is approximately 400 nm, which is significantly shorter than the dimensions of the single crystal sample (2.5 $\times$ 2.5 $\times$ 6 mm$^3$) used to measure $\kappa$. 
We also examined the phonon-isotope effects~\cite{lindsay2013phonon} by considering different isotope masses in HNEMD simulations, and find that these effects are negligible for $\kappa$.

In summary, we synthesized a single-crystal inorganic halide perovskite Cs$_3$Bi$_2$I$_6$Cl$_3$ and measure its glassy $\kappa$ from 20~K to 300~K. 
We developed a machine-learning neuroevolution potential and performed PIMD simulations to investigate its atomic structure. 
The atomic probability distribution from PIMD trajectories reveal an disorder-to-order transition at $\sim$200~K, indicating significant lattice distortion in Cs$_3$Bi$_2$I$_6$Cl$_3$. 
Given the disordered structure in Cs$_3$Bi$_2$I$_6$Cl$_3$ at low temperatures, we employed both lattice dynamics and molecular dynamics approaches to compute its $\kappa$. 
We concluded that evaluating the glass limit of the WTE in our structurally distorted atomistic models of Cs$_3$Bi$_2$I$_6$Cl$_3$ (i.e., a calculation at $\mathbf{q}=\mathbf{0}$ only in a model containing more than four thousands  atoms) rationalizes the glassy $\kappa$ below 200~K. Moreover, the nearly temperature-independent conductivity of the ordered crystal Cs$_3$Bi$_2$I$_6$Cl$_3$ at and above 250~K originates from strong wave-like tunneling transport, as explained by the solution of the WTE in the complex crystal regime (i.e., a calculation involving an integration of phonon properties over the entire BZ).
Our $\kappa$ computation workflow considers both nuclear quantum effects and atomic disorder, providing a framework for comprehending and modeling heat transport in disordered materials.

\section{Methods}

\subsection{Experiments}
\paragraph{Sample preparation}
The raw materials CsCl (99.9\%, Innochem) and BiI$_3$ (99.99\%, Aladdin) were sealed in a quartz tube and placed in a two-zone vertical furnace. The tube was gradually heated to 1023 K, where it was maintained for 40 hours. Subsequently, the lower zone of the furnace was cooled to 903 K over an 8-hour period. The temperatures in both zones were then decreased at a rate of 3 K/h for a total of 90 hours. Finally, the heating was turned off to complete the cooling process, allowing the resulting Cs$_3$Bi$_2$I$_6$Cl$_3$ to crystallize. 

\paragraph{Sample characterization}
Single-crystal X-ray diffraction (XRD) studies were performed on a Bruker D8 QUEST diffractometer equipped with a Mo K$\alpha$ radiation source ($\lambda$ = 0.71073 Å) at room temperature. The crystal structure solution and refinements of Cs$_3$Bi$_2$I$_6$Cl$_3$ is carried out using single crystal XRD at room temperature, as shown in Table S2. The Cs$_3$Bi$_2$I$_6$Cl$_3$ sample possesses the trigonal structure with a space group $P\overline{3}m1$. As presented in Figure S1, all the peaks of powder XRD pattern fit well with the Cs$_3$Bi$_2$I$_6$Cl$_3$ pattern. The XRD pattern tested on the cleavage plane shows that the crystal is liable to fracture along the $(00l)$ crystal plane due to the weak chemical bonds between the layers. 

\paragraph{Thermal conductivity measurement}
The cleavage plane of crystalline Cs$_3$Bi$_2$I$_6$Cl$_3$ is the $(00l)$ ($l$ is an integer) plane. To measure the $\kappa$ in the cleavage plane ($\kappa_{x}$ or $\kappa_{y}$) and across the cleavage plane ($\kappa_{z}$), two cubes (sectional area: 2.5 mm $\times$ 2.5 mm; height: 6 mm) were cut from the obtained crystal for the thermal conductivity measurement. The low-temperature $\kappa$ was measured using a physical properties measurement system (PPMS, Quantum Design). The Cs$_3$Bi$_2$I$_6$Cl$_3$ crystal is nonconductive \cite{acharyya2022glassy}, so the measured thermal conductivity is equal to the lattice thermal conductivity. Our measurements of $\kappa$ agree very well with the previous experiments \cite{acharyya2022glassy}.

\subsection{Calculations}

\paragraph{Constructing the MLP.}
We utilized the neural-network-based neuroevolution potential (NEP) approach \cite{fan2021neuroevolution} to develop a MLP for Cs$_3$Bi$_2$I$_6$Cl$_3$. The procedure begins with obtaining the relaxed structure at 0~K using the VASP package \cite{kresse1996efficient, kresse1999ultrasoft}. Subsequently, we conducted \textit{ab initio} molecular dynamics (AIMD) simulations in the NVT ensemble with a 2 $\times$ 2 $\times$ 2 supercell (112~atoms) at temperatures of 20~K, 50~K, 100~K, 200~K, 300~K, 400~K, 500~K, and 600~K, randomly selecting $\sim$50 configurations for training at each temperature.
Furthermore, AIMD simulations with the NPT ensemble, varying pressure ($\pm$ 1~Gpa and 2.5~Gpa) at specific temperatures (25~K, 150~K and 300~K), were performed and $\sim$ 30 configurations were randomly chosen for each condition to account for lattice thermal expansion. Taking into account the two-dimensional layered structure of Cs$_3$Bi$_2$I$_6$Cl$_3$, we further selected configurations using the initial NEP, conducting MD simulations with D3 dispersion correction \cite{ying2023combining} at 
25~K, 100~K, 200~K, and 300~K. The final training set comprises 1000 configurations.
We used the PBEsol functional \cite{perdew2008restoring} in the DFT calculations, known for its accuracy in predicting lattice constants and $\kappa$ in strongly anharmonic crystals \cite{wei2024hierarchy}.
All AIMD simulations employ an energy cutoff value of 400~eV and a $\Gamma$-centered 1 $\times$ 1 $\times$ 1 $k$-point mesh with the PBEsol \cite{perdew2008restoring} functional. 
We raised the energy cutoff to 550~eV with a denser 3 $\times$ 3 $\times$ 3 $k$-point mesh to perform accurate single-point DFT calculations and obtain the energies and atomic forces of the selected configurations. We used the GPUMD package \cite{fan2017efficient} to train the NEP model, and the final training/test root mean square errors (RMSEs) for energy, atomic force, and virial are 0.35/0.33~meV/atom, 0.027/0.026~eV/Å, and 3.79/3.33~meV/atom, respectively. 

\paragraph{Cross validation of the NEP model.}
To validate our NEP model, we performed path integral molecular dynamics (PIMD) simulations at selected temperatures (25~K, 100~K, 300~K, and 400~K), both with and without the D3 correction, and sampled a number of configurations for subsequent single-point DFT calculations. Parity plots between NEP and DFT for the energies and forces of the sampled configurations are presented in Figs. S2 and S3. Without the D3 correction, The RMSEs for energy and force are 0.36~meV/atom and 0.025~eV/Å, respectively, while they are 0.31~meV/atom and 0.021~eV/Å, respectively, with the D3 correction.
The lattice constants (Table S1) and atomic potential energy surfaces of some soft modes (Fig. S12) from DFT and NEP calculations are in good agreement.

\paragraph{MD simulations}
We performed MD simulations using the GPUMD package \cite{fan2017efficient}, for which PIMD techniques have been implemented recently \cite{ying2024arxiv}. The integration time step was set to 1~fs in both classical and PIMD related simulations. 
PIMD/RPMD simulations were conducted at 25~K and 50~K with 64~beads, at 100~K, 150~K, and 200~K with 32~beads, and at 250~K and 300~K with 16~beads. Convergence tests for the number of beads in these simulations are provided in Fig. S14.
For quantum MD simulations, we started with PIMD simulations in an NPT ensemble to obtain equilibrium structures at the specified temperatures. 
Following this, we run 0.5~ns RPMD simulations to collect atomic trajectories to calculate atomic probability distributions and radial distribution functions, or TRPMD simulations to compute $\kappa$ based on the Green-Kubo relationship. 
Further details on the computation of $\kappa$ using Green-Kubo theory with TRPMD simulations are available in Fig. S9 and Note S1 of SI. 
For classical HNEMD simulations, we used an NPT ensemble, employing a stochastic cell rescaling barostat \cite{bernetti2020pressure} in combination with the stochastic velocity rescaling thermostat \cite{bussi2007canonical} to calculate $\kappa$. 
We conducted a thorough examination of size effects and the selection of external forces during HNEMD simulations.

\paragraph{Anharmonic linewidths and its analytical function}

Calculating vibrational linewidths rigorously in a complex, atom-heavy primitive cell is highly time-consuming. 
To address this challenge, previous studies \cite{simoncelli2023thermal,thebaud2022perturbation,harper2024vibrational} introduced a coarse-grained function, $\Gamma_a[\omega]$, in various formats to reduce computational demands. The only required input to obtain $\Gamma_a[\omega]$ is the exact linewidths at $\textbf{q}$ = 0.  Simoncelli \textit{et al.} \cite{simoncelli2023thermal} demonstrated that the use of the single-valued function $\Gamma_a[\omega]$ leads to practically identical results for the conductivity of rWTE compared to exact linewidths. This fitting was demonstrated to reliably capture the linewidths of $v$-SiO$_2$ \cite{simoncelli2023thermal}, Mg$_2$Si$_{1-x}$Sn$_{x}$ \cite{thebaud2022perturbation}, amorphous Si \cite{fiorentino2023hydrodynamic} and Al$_2$O$_3$ \cite{harper2024vibrational}. 

Here we used the RPMD simulations with normal mode decomposition method \cite{mcgaughey2014predicting,carreras2017dynaphopy} to obtain the exact anharmonic linewidths of all vibrational modes at $\textbf{q}$ = 0, and then we fit \cite{simoncelli2023thermal} the $\Gamma_a[\omega]$ as
\begin{equation}
    \Gamma_{\mathrm{a}}[\omega]=\frac{1}{\sqrt{\frac{1}{\left(\Gamma_1[\omega]\right)^2}+\frac{1}{\left(\Gamma_2[\omega]\right)^2}}},
\end{equation}
where $\Gamma_1[\omega]$ and $\Gamma_2[\omega]$ are defined as
\begin{equation}
    \Gamma_1[\omega]=\frac{\sum_{\mathbf{q}=\mathbf{0}, s} \frac{1}{\sqrt{2 \pi \sigma^2}} \exp \left[-\frac{\hbar^2\left(\omega(\mathbf{q})_s-\omega\right)^2}{2 \sigma^2}\right]}{\sum_{\mathbf{q}=\mathbf{0}, s} \frac{1}{\Gamma(\mathbf{q})_s \sqrt{2 \pi \sigma^2}} \exp \left[-\frac{\hbar^2\left(\omega(\mathbf{q})_s-\omega\right)^2}{2 \sigma^2}\right]},
\end{equation}

\begin{equation}
    \begin{aligned}
& \Gamma_2[\omega]=p \cdot \omega^2 \text {, } \\
& p=\frac{\sum_{\mathbf{q}=0, s} \int_{\omega_0}^{2 \omega_0} d \omega_c \frac{\Gamma(\mathbf{q})_s}{\omega^2(\mathbf{q})_s \sqrt{2 \pi \sigma^2}} \exp \left[-\frac{\hbar^2\left(\omega(\mathbf{q})_s-\omega_c\right)^2}{2 \sigma^2}\right]}{\sum_{\mathbf{q}=0, s} \int_{\omega_0}^{2 \omega_0} d \omega_c \frac{1}{\sqrt{2 \pi \sigma^2}} \exp \left[-\frac{\hbar^2\left(\omega(\mathbf{q})_s-\omega_c\right)^2}{2 \sigma^2}\right]}. \\
&
\end{aligned}
\end{equation}
Here, $\omega_0$ is the smallest non-zero frequency when $\mathbf{q} = 0$, and $\sigma$ = 70 cm$^{-1}$ is a broadening parameter set large enough to ensure smooth averaging of linewidths. 
The full-order lattice anharmonicity is intrinsically considered for anharmonic linewidths extracted from the normal mode decomposition with ring-polymer or classical MD simulations (see Fig. S5 of SI for comparisons).

\paragraph{WTE thermal conductivity of disordered Cs$_3$Bi$_2$I$_6$Cl$_3$} 
At low temperatures (25~K to 200~K), RPMD simulations were performed for supercells of different sizes with \(P\overline{3}m1\) structure at various temperatures. 
A configuration randomly extracted from the equilibrium stage was then relaxed,
and the resulting structures were used as the initial distorted configurations. 
Harmonic force constants were computed using the finite displacement method implemented in the Phonopy package \cite{togo2015first,togo2023implementation,togo2023first}. 
Frequencies and group velocity matrices for vibrational modes at $\mathbf{q=0}$ were derived from the harmonic force constants using the Phono3py package \cite{togo2015first,togo2023implementation,togo2023first}.
The physical linewidths were obtained from normal mode decomposition with RPMD trajectories, and analytical linewidths were fitted as described above. Figure S12 provides a detailed workflow for computing WTE $\kappa$ at $\mathbf{q=0}$ for disordered materials.

\paragraph{WTE thermal conductivity of crystal Cs$_3$Bi$_2$I$_6$Cl$_3$}
We obtained the temperature-dependent cubic and quartic interatomic force constants (IFCs) at 250~K and 300~K using the hiPhive package \cite{eriksson2019hiphive}. 
The harmonic terms at 0~K calculated using Phonopy \cite{togo2015first} were subtracted from the atomic forces and only cubic and quartic IFCs were extracted to the residual force-displacement data. We collected the force-displacement data from RPMD simulations.
Three-phonon linewidths were calculated using the ShengBTE \cite{li2014shengbte} package. Four-phonon linewidths were calculated using our code based on the formulae developed by Feng \textit{et al.} 
\cite{feng2016quantum,feng2017four}.  
We carefully tested the relation between $q$-point mesh and $\kappa$, and a 6 $\times$ 6 $\times$ 6 $q$-point mesh was used to compute the phonon linewidths and the $\kappa$ of Cs$_3$Bi$_2$I$_6$Cl$_3$. 
The $\texttt{scalebroad}$ parameter for Gaussian smearing used for numerically ensuring the energy conservation process in phonon-phonon scatterings was set to 1 and 0.1 for the calculations of the three- and four-phonon scatterings, respectively. 
We also calculated the WTE conductivity based on our experimental lattice constants at 300~K, and the calculated WTE $\kappa$ ($\kappa_x$ = 0.227~Wm$^{-1}$K$^{-1}$; $\kappa_z$ = 0.231~Wm$^{-1}$K$^{-1}$) based on experimental lattice constants show a better agreement with the measurements, compared to those based on the predicted lattice constants of NEP.

\paragraph{HNEMD and its quantum correction}
In HNEMD \cite{fan2019homogeneous,evans1982homogeneous}, one applies an external driving force 
\begin{equation}
    \boldsymbol{F}_i^{\mathrm{ext}}=\boldsymbol{F}_{\mathrm{e}} \cdot \mathbf{W}_i
\end{equation}
to generate a nonzero heat current. Here $\boldsymbol{F}_{\mathrm{e}}$ is the driving force parameter and $\mathbf{W}_i$ is the virial tensor for atom $i$.
Within the linear-response regime, the heat current $\mathbf{J}$ is proportional to the driving force parameter as
\begin{equation}
    \left\langle J^\alpha\right\rangle=T V \sum_\beta \kappa^{\alpha \beta} F_{\mathrm{e}}^\beta,
\end{equation}
where $V$ is the volume of the simulated system, and $\alpha$ and $\beta$ are Cartesian directions. 
Considering diagonal elements of the thermal conductivity tensor and dropping the tensor indices, the cumulative thermal conductivity in one direction can be expressed as 
\begin{equation}
    \kappa(t)=\frac{1}{t} \int_0^t d s \frac{\left\langle J(s)\right\rangle}{T V F_{\mathrm{e}}}.
\end{equation}
The HNEMD approach \cite{fan2019homogeneous} also allows for obtaining the spectral thermal conductivity
\begin{equation}
    \kappa(\omega, T)=\frac{2}{V T F_{\mathrm{e}}} \int_{-\infty}^{\infty} d t e^{i \omega t} K(t),
\end{equation}
Here, $K(t)$ is the virial-velocity correlation function and its vectorial version is defined as 
\begin{equation}
    \boldsymbol{K}(t)=\sum_i\left\langle\mathbf{W}_i(0) \cdot \boldsymbol{v}_i(t)\right\rangle. 
\end{equation}
The classical spectral thermal conductivity $\kappa(\omega, T)$ can be quantum-corrected \cite{turney2009assessing,wang2023quantum} by multiplying it with the ratio of quantum-to-classical modal heat capacity:
\begin{equation}
    \kappa^{\mathrm{q}}(\omega, T)=\kappa(\omega, T) \frac{x^2 e^x}{\left(e^x-1\right)^2},
\end{equation}
where $x = \hbar \omega / (k_B T)$.

\paragraph{Spectral Mean free path}
To compute the vibrational mean free paths (MFPs) of the vibrational modes, we first perform a single NEMD simulation in the ballistic limit (low \(T\) and short supercell length \(L\)), equivalent to the atomistic Green's function approach, and then employ the same spectral decomposition method as in HNEMD to obtain the spectral thermal conductance $G(\omega)$ \cite{wang2023quantum}:
\begin{equation}
    G(\omega) = \frac{2}{V \Delta T} \int_{-\infty}^{\infty} dt \, e^{i \omega t} K(t),
\end{equation}
where $\Delta T$ is the temperature difference between the heat source and the heat sink in the NEMD setup. Then we can derive the spectral MFP as \cite{fan2019homogeneous}
\begin{equation}
    \lambda(\omega, T) = \kappa(\omega, T) / G(\omega).
\end{equation}

\textbf{Acknowledgments}
ZZ acknowledges the European Union’s Horizon 2020 research and innovation programme under the Marie Skłodowska-Curie grant agreement No 101034413.
The authors acknowledge the research computing facilities offered by HPC ISTA and ITS HKU.

\nocite{*}
\bibliography{mainrefs}
\end{document}